\documentclass{elsart}
\usepackage{epsfig}
\usepackage{bm}
\usepackage{graphicx}

\begin{document}

%
\hbox to \textwidth{
\lower 2.0cm
\hbox to 3.8cm{
\hss
\epsfysize3cm
\epsfbox{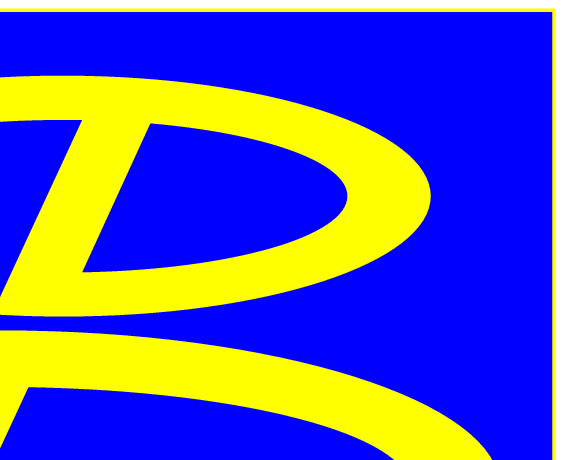}    
}

\hss
\hbox to 3cm{
\begin{tabular}{r}
{KEK preprint 2002-109} \\
{Belle preprint 2002-34} \\
{UCTP-106-02}
\end{tabular}
\hss
}
}

\vspace{12pt}

\title{\boldmath Study of $\bar{B^{0}} \to D^{(*)0} \pi^+ \pi^-$ Decays}

\begin{frontmatter}
\date{\today}

\begin{abstract}
We report on a study of $\bar{B^{0}} \to D^{(*) 0} \pi^+ \pi^-$ decays
using 29.1 fb$^{-1}$ of $e^{+}e^{-}$ annihilation data recorded 
at the $\Upsilon(4S)$ resonance with the Belle detector at the KEKB storage ring.
Making no assumptions about the intermediate mechanism,
the branching fractions for  $\bar{B}^0 \to D^0 \pi^+ \pi^-$ 
and $\bar{B}^0 \to D^{* 0} \pi^+ \pi^-$ are determined
to be $(8.0 \pm 0.6 \pm 1.5) \times 10^{-4} $ and 
$ (6.2 \pm 1.2 \pm 1.8) \times 10^{-4}$
respectively. An analysis of $\bar{B^{0}} \to D^{0} \pi^+ \pi^-$
candidates yields to the first observation of 
the color-suppressed hadronic decay $\bar{B}^0 \to D^0 \rho^0$  
with the branching fraction
$(2.9 \pm 1.0 \pm 0.4) \times 10^{-4}$. 
We measure the ratio of branching fractions 
${\mathcal B}(\bar{B^0} \to D^0 \rho^0)$/${\mathcal B}(\bar{B^0} \to D^0 \omega)$ 
= 1.6 $\pm$ 0.8.

\vspace{3\parskip}
\end{abstract}

\end{frontmatter}

{\renewcommand{\thefootnote}{\fnsymbol{footnote}}
\normalsize

\collab{The Belle Collaboration}
  \author[KEK,Cincinnati]{A.~Satpathy}, 
  \author[KEK]{K.~Abe}, 
  \author[Niigata]{R.~Abe}, 
  \author[Tohoku]{T.~Abe}, 
  \author[KEK]{I.~Adachi}, 
  \author[Tokyo]{H.~Aihara}, 
  \author[Nagoya]{M.~Akatsu}, 
  \author[Tsukuba]{Y.~Asano}, 
  \author[Toyama]{T.~Aso}, 
  \author[ITEP]{T.~Aushev}, 
% ...

% ...
\address[Cincinnati]{University of Cincinnati, Cincinnati, OH, USA}
% ...
\address[KEK]{High Energy Accelerator Research Organization (KEK), Tsukuba,
Japan}
% ...
\address[ITEP]{Institute for Theoretical and Experimental Physics, Moscow,
Russia}
% ...
\address[Nagoya]{Nagoya University, Nagoya, Japan}
% ...
\address[Niigata]{Niigata University, Niigata, Japan}
% ...
\address[Tohoku]{Tohoku University, Sendai, Japan}
\address[Tokyo]{University of Tokyo, Tokyo, Japan}
% ...
\address[Toyama]{Toyama National College of Maritime Technology, Toyama, Japan}
\address[Tsukuba]{University of Tsukuba, Tsukuba, Japan}
% ...
\thanks[NovaGorica]{on leave from Nova Gorica Polytechnic, Nova Gorica, Slovenia}


\normalsize 
\section{Introduction}

Exclusive hadronic decay rates 
provide important tests of models for $B$ meson decay \cite{Fakirov}.
$B$ decays to final states that include a $D^0$ or a $D^{*0}$
accompanied by two charged pions are interesting, because 
such decays provide a precision testing ground 
for factorization \cite{Isgur},
and because one can search for resonant substructure in the 
final state. At present, only an upper limit 
${\mathcal B}(\bar{B^{0}} \to D^0  \pi^+ \pi^-) < 1.6 \times 10^{-3}$ \cite{cleo_alam},
has been measured.
The $D^{(*)0} \pi^+ \pi^-$ final state includes the 
$\bar{B^{0}} \to D^{(*)0} \rho^{0}$ decay which has not
yet been observed~\cite{D0X0}. As shown in Fig.~1(a)
this decay proceeds via an internal spectator diagram,
and is ``color-suppressed'' since the color of the quarks
produced by the weak current must correspond to the color
of the $c$ quark and the spectator quark.
Recent measurements \cite{col_supp} of the branching fractions
for the color-suppressed decays $\bar{B^0} \to D^{0} X^{0}$, 
where $X^0 = \pi^0, \eta $ or $\omega$,
are all higher than theoretical
predictions \cite{Neubert1} providing evidence for failure of 
the na\"{\i}ve factorization model and suggesting sizable
final state interactions (FSI). In the heavy quark limit, 
the QCD factorization model works effectively for color-allowed decays, 
while color-suppressed decays require substantial correction \cite{Neubert2}.
Assuming SU(3) symmetry for the FSI rescattering phase,
the observed discrepancy can be accommodated and
branching fractions, such as ${\mathcal B}(\bar B^0\to D^0\rho^0)$, 
can be predicted \cite{George}.
It is important to test whether  $\bar{B^0} \to D^{0} \rho^0$, once observed, supports
the current pattern of QCD effects in color-suppressed B decays.

\begin{figure}[h]\centering
\mbox{\psfig{file=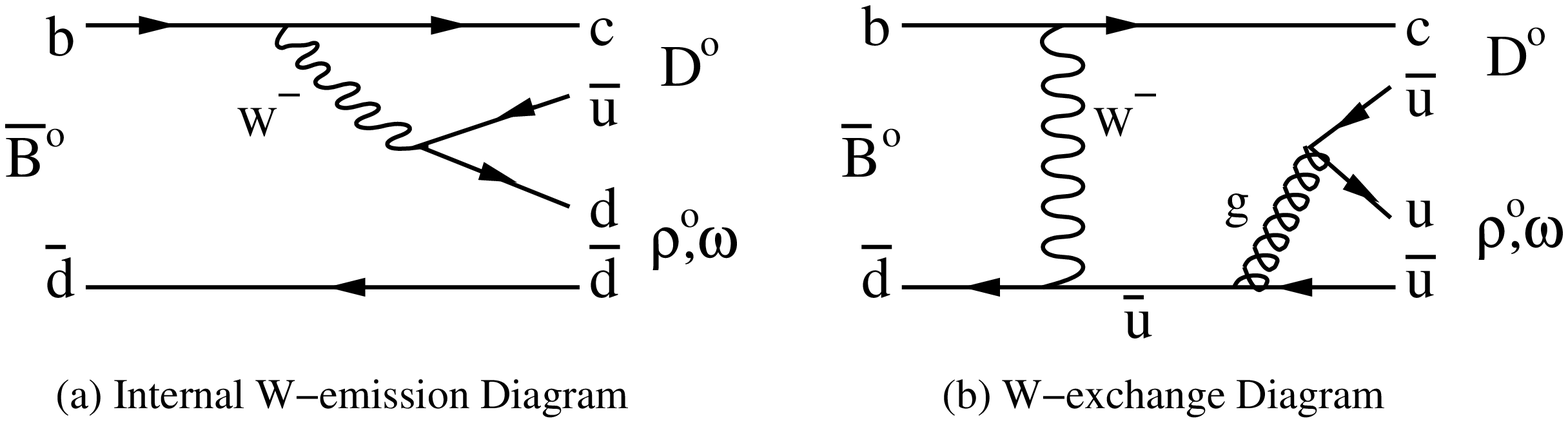,height=1.7in, width=4.5in}}
\label{faynmandiag_fig1}
\centering 
\vspace{0.5cm}
\caption{ 
Decay diagrams for $\bar{B^0} \to D^0 \rho^0, D^0 \omega$.}
\end{figure}

The dominant diagrams for such neutral $B$ meson decays preserve the
spectator $d$-quark and therefore require that the final state 
neutral light meson be produced via its $d-\bar{d}$ component (Fig.~1(a)).
These diagrams predict equal branching fractions
for $D^0 \rho^0$ and $D^0 \omega$ and for $D^{*0} \rho^0$ 
and $D^{*0} \omega$. Other diagrams, such as W-exchange (Fig.~1(b)) or final state 
interactions can produce the $u-\bar{u}$ state and   
therefore give different branching fractions.
This equality is therefore a very sensitive test
for small amplitudes in which the spectator $d$-quark does
not appear in the final state and the $\rho$ or $\omega$ are 
produced via their $u-\bar{u}$ components \cite{George,Lipkin}. 

In this paper, we will report on the study of 
$\bar{B^0}$ decays that have one 
$D^0$ or $D^{*0}$ and two oppositely charged pions in 
the final state. Inclusion of charge conjugate modes
is implied throughout this paper.

\section{Data Sample and Selection Criteria}

The data sample used in this paper was collected with the
Belle detector at KEKB \cite{KEKB}. It is based on an 
integrated luminosity of 29.1 fb$^{-1}$ at the $\Upsilon(4S)$ resonance,
corresponding to 31.3 million $B \bar{B}$ events.

The Belle detector~\cite{Belle} is a large-solid-angle magnetic
spectrometer that
consists of a three-layer silicon vertex detector,
a 50-layer central drift chamber (CDC), an array of
aerogel threshold \v{C}erenkov counters (ACC), 
a barrel-like arrangement of time-of-flight
scintillation counters (TOF), and an electromagnetic calorimeter
comprised of CsI(Tl) crystals (ECL)  located inside 
a super-conducting solenoid coil that provides a 1.5~T
magnetic field.  An iron flux-return located outside of
the coil is instrumented to detect $K_L^0$ mesons and to identify
muons (KLM). 

Hadronic event selection is described elsewhere~\cite{HadronB}.
$\pi^{0}$ candidates are formed by 
combining two photons detected in the ECL, whose
invariant mass is within a $\pm 16$ MeV/$c^2$ mass window 
around the $\pi^{0}$ peak. The $\pi^0$ daughter photons
are required to have energies greater than 20 MeV.
We require the point of closest approach to the origin of each track
to be within $\pm 5$ mm from the beam axis and $\pm 3$ cm along the beam axis 
from the interaction point to remove background.
Tracks identified as electrons (from the responses of the CDC
and ECL) or muons (from the response of the KLM)
are removed. Kaon and pion candidates are distinguished by 
combining the $dE/dx$ information from the 
CDC, time of flight information from the TOF and hit information
from the ACC.

$D^{0}$ candidates are reconstructed 
in the decay modes $K^{-} \pi^{+}$, $K^{-} \pi^{+} \pi^{0}$, and 
$K^{-} \pi^{+} \pi^{-} \pi^{+}$.
For $D^0 \to K^- \pi^+ \pi^0$, the $\pi^0$ daughter photons
are required to have energies greater than 50 MeV and
we select regions of the Dalitz plot with large
decay amplitudes to further suppress the combinatorial 
background~\cite{D02kpipi0}.
The invariant masses of $D^{0}$ candidates are required to be 
within $2.5 \sigma $ of the nominal mass.
The selected $\pi^0$s and $D^0$s are then kinematically
fit with their masses constrained to their nominal values \cite{PDG2002}.
$D^{*0}$ candidates are formed by combining $D^0$ and $\pi^0$ 
candidates and selecting those with 
mass difference $\delta m = M_{D^{*0}} - M_{D^{0}}$ in the 
range 0.1400 GeV/$c^2 < \delta m <$ 0.1445 GeV/$c^{2}$.

\section{$B$ Meson Reconstruction}

After selecting $D^0$ and $D^{*0}$ candidates, 
we combine them with two oppositely charged
pions to form $B$ candidates. 
The two oppositely charged candidate pions from the $B$ decay 
are required to come from a single vertex. 
To remove $K^0_S$ candidates from the sample,
the distance of the $\pi^{+}\pi^{-}$
vertex from the beam interaction point in the $r-\phi$ 
plane is required to be less than 0.8~cm.
Two kinematic variables are used to identify
signal candidates, the beam constrained mass, 
$M_{bc} = \sqrt{(E^{CM}_{beam})^2 - (P^{CM}_B)^{2}}$, 
and the energy difference $\Delta E = E^{CM}_B - E^{CM}_{beam}$
where $E^{CM}_{B}$ and $P^{CM}_B$ are
the center of mass (CM) energy and momentum of the $\bar{B}^0$ candidate,
and $E^{CM}_{beam}$ = $\sqrt{s}/2 = 5.29$ GeV.
We select events with $|\Delta E|< 0.2$~GeV 
and 5.272~GeV/$c^2$ $<M_{bc}<$ 
5.288~GeV/$c^2$ (5.271~GeV/$c^2$ $<M_{bc}<$ 5.289~GeV/$c^2$)
for $D^0 \pi^+ \pi^-$ ($D^{*0} \pi^+ \pi^-$) final states. 
Further, if there are multiple $B$ candidates in an event, we choose the 
candidate with the smallest $\chi^2$ combination, 
\begin{equation}
\chi^{2} = \chi^2_{D^0} + \chi^2_{\pi^+ \pi^-} (+ \chi^2_{\delta m})
\end{equation}
where, $\chi^2_{D^{0}}$ and $\chi^2_{\pi^+ \pi^-}$ are obtained from 
$D^0$ and $\pi^+ \pi^-$ vertex fitting respectively.  
For decay modes containing $D^{*0}$, $\chi^2_{\delta m}$---  
defined as the square of the difference of $\delta m$
from its nominal value, in units of its resolution,
$(\Delta (\delta m) / \sigma (\delta m))^2$---
is additionally included in the best candidate selection requirement.

\section{Background Suppression}

Since the continuum background 
(arising from $e^{+}e^{-} \rightarrow q\bar{q}
(q=u,d,c,s)$ transitions) has a different event topology,
shape variables are very effective at improving the 
signal to noise ratio. Events are required to
satisfy $R_2 < 0.5$, where $R_2$ is the ratio of the second Fox-Wolfram moment
to the zeroth moment determined using charged tracks and 
unmatched neutral showers
\cite{Fox}. The angle between the $B$ candidate direction
and the thrust axis  \cite{Farhi} of the rest of the event ($\theta_{T}$) is 
required to satisfy $|\cos(\theta_{T})| < 0.7$.

For the $\bar{B}^0 \to D^{(*)0} \pi^+ \pi^-$ branching fraction
measurements, we make no assumptions about the intermediate mechanism,
except that we reject the large contribution from the well-established
decay $\bar{B}^0 \to D^{*+} \pi^-$ to the $D^0 \pi^+ \pi^-$ final
state.  These events are rejected by requiring 
$M^2_{D^0\pi^+} > 4.62\,\mathrm{GeV}^2/c^4$ (Fig. 2), 
which removes 1\% of the phase space for
$\bar{B}^0 \to D^{0} \pi^+ \pi^-$.  
As the decay $\bar{B^0} \to D_2^{*}(2460)^{+} \pi^-$
is not well established~\cite{PDG2002}, no attempt is made 
to reject it and this mode
is thus included in our branching fraction measurement.

Color-favored decays can also cause a 
background when a final state pion is replaced by
a pion from the decay of the other $B$ (for example
$B^- \to D^{(*)0} \rho^-$ may be reconstructed as 
$\bar{B^0} \to D^{(*)0} \pi^+ \pi^-$). To reduce this
background we veto events which can also be reconstructed
in a color-favored mode. This requirement removes 1\% of the signal
candidates. Using a sample of 44 million generic $b \to c$ 
decays generated via Monte Carlo (MC) simulation, 
the small remaining background is studied and found
not to peak in $M_{bc}$ or $\Delta E$.

\section{Branching Fractions for 
$D^0 \pi^+ \pi^-$ and $D^{*0} \pi^+ \pi^-$ final states}

The distribution in $\Delta E $ for 
the surviving candidates for $\bar{B}^0 \to D^0 \pi^+ \pi^-$
is shown in Fig.~3(a).
Since intermediate resonances dominate the decay rate
we obtain a non-uniform distribution of events 
on the Dalitz plot. In addition, the 
efficiency varies across the Dalitz plot due to momentum
dependences of the reconstruction and particle identification efficiencies.
We divide the Dalitz plot into six different regions 
expected to be dominated by different intermediate processes
as shown in Fig.~2
and determine the efficiency \cite{efficiency} and signal yield (from $\Delta E$ fit) 
for each. Table~\ref{dpipiBF} summarizes our results.

\begin{figure}[h]
\begin{center}
\mbox{\epsfig{file=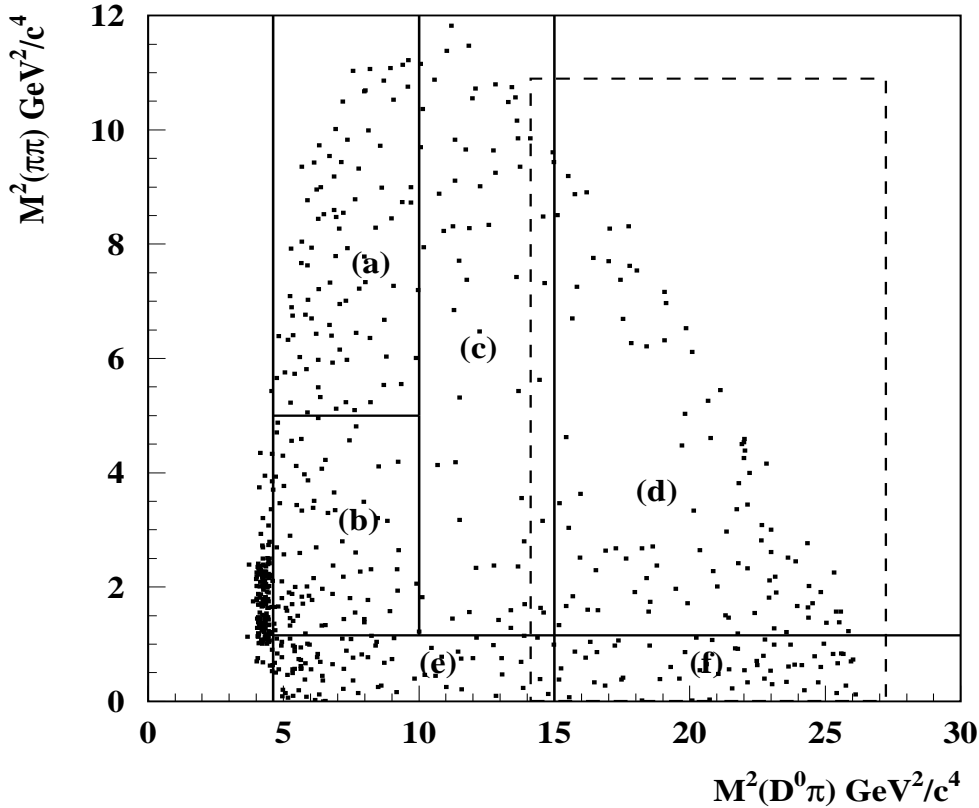,height=5.0in, width=5.5in}
}
\label{dalitz_plot}
\vspace{0.5cm}
\caption{
Dalitz plot for $\bar{B}^0 \to D^0 \pi^+ \pi^-$ events with $|\Delta E|<0.03$ GeV,
showing the regions (a)-(f) used in the efficiency measurement.
Events in the dashed box 
are used for the branching fraction measurement of  
$\bar{B}^0 \to D^0 \rho^0$ as
explained in the text.}
\end{center}
\end{figure}

\begin{figure}
\centering
\hbox{\epsfig{file=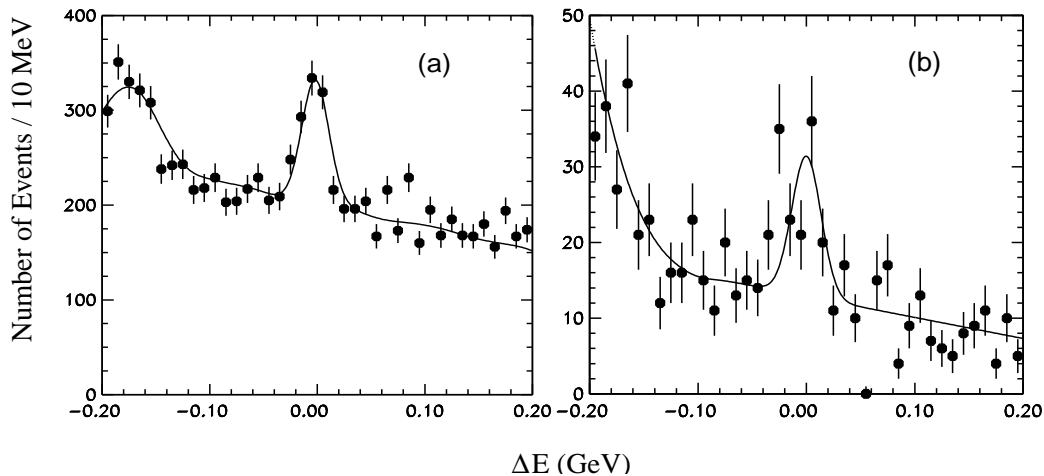,height=2.7in, width=5.5in}
}
\label{deltaE_fig3}
\centering
\caption{
(a) $\Delta E$ distribution for $\bar{B} \to D^{0} \pi^+ \pi^-$
events satisfying $M_{D^0 \pi^+}^2 >$ 4.62 GeV$^2/c^{4}$. 
(b) $\Delta E$ distribution for 
$\bar{B} \to D^{*0} \pi^+ \pi^-$ candidates
with no requirement on $M_{D^{*0} \pi^+}^2$. }
\end{figure}

\begin{table}[htb]
\caption{Summary of branching fraction results for $\bar{B^0} \to D^{0} \pi^+ \pi^-$
in different regions of the Dalitz plot. 
The last row gives the sums of the signal yields and branching fractions.}
\label{dpipiBF}
\begin{tabular}{c c c c} 
\hline \hline
Region & Efficiency (\%) & Signal Yield & Branching Fraction ($\times 10^{-4}$) \\
\hline
(a) & 1.87 $\pm$ 0.09 & 98 $\pm$ 15  &1.7 $\pm$ 0.3 \\
(b) & 1.66 $\pm$ 0.11 & 70 $\pm$ 18  &1.3 $\pm$ 0.3 \\
(c) & 1.88 $\pm$ 0.08 & 17 $\pm$ 5   &0.3 $\pm$ 0.1 \\
(d) & 1.94 $\pm$ 0.07 & 57 $\pm$ 15  &0.9 $\pm$ 0.2 \\
(e) & 2.10 $\pm$ 0.17 & 76 $\pm$ 19  &1.2 $\pm$ 0.3 \\
(f) & 1.85 $\pm$ 0.12 & 150 $\pm$ 16  &2.6 $\pm$ 0.3 \\ \hline
Total &               & 469 $\pm$ 38  &8.0 $\pm$ 0.6 \\ \hline \hline

\end{tabular}
\end{table}

For each Dalitz plot region
we model the signal in $\Delta E$ with a Gaussian
function where both the mean and width
are fixed from MC studies. The background shape 
in this fit is modeled by two components: (1) a linear shape 
for continuum background obtained from the sideband data 
(5.20~GeV/$c^2 < M_{bc} <5.26$~GeV/$c^2$); 
(2) a smooth histogram shape for $\bar{B^0} \to D^{*0} \pi^+ \pi^-$
feed-down obtained from MC. The normalizations of the signal
and background components are free parameters in the fit.
We obtain the branching fraction
for $\bar{B^0} \to D^{0} \pi^+ \pi^-$ by taking the sum of the
branching fractions in the six regions of the Dalitz plot
and making a correction of 1\% for the unobserved 
region where  $M_{D^0 \pi^+}^2 < 4.62\,\mathrm{GeV}^2/c^4$.
In all branching fraction
calculations we assume equal production of $B^0\bar{B}^0$ and $B^+B^-$
pairs from the $\Upsilon(4S)$.

To estimate the branching fraction for $\bar{B^0} \to D^{*0} \pi^+ \pi^-$ 
decays, we make no restriction on $M_{D^{*0} \pi^+}^2$.
Due to limited statistics, we do not estimate the branching fraction
region by region. Instead, we use the yield from the $\Delta E$ fit (Fig.~3(b))
and include a model dependent systematic error (19\%) that arises from the
difference between the detection efficiency when 
the signal MC events are $\bar{B^0} \to D^{*0} \pi^+ \pi^-$
and $\bar{B^0} \to D^{*0} \rho^0$. 
The two detection efficiencies
are 0.26\% and 0.32\%, respectively where 
the $\bar{B}^0 \to D^{*0} \rho^0$ decay is generated with equal
rates to each helicity state.

The background near
the lower side of the $\Delta E $ distribution is modeled by
$B^+ \to D^{*0} a_{1}^{+}$ feed-down measured using MC.
The yield from the fit is 62 $\pm$ 12 events.
We measure the branching fraction for $\bar{B^0} \to D^{*0} \pi^+ \pi^-$ 
using the phase space MC efficiency. 
The results are summarized in Table 2.

\section{Search for Color-Suppressed $\bar{B^0} \to D^{(*)0} \rho^0$ Decays}

Multi-body decays of $B$ mesons can occur through various strong resonances
that can interfere with each other. We search for color-suppressed 
$\bar{B^0} \to D^{(*)0} \rho^0$ decays in the $D^{(*)0} \pi^+ \pi^-$
final state. We study the $\pi^+ \pi^-$ invariant mass of the
events in the signal region ($|\Delta E|<0.030$ GeV for 
$D^0 \pi^+ \pi^-$ and $|\Delta E|<0.035$ GeV for 
$D^{*0} \pi^+ \pi^-$) and 
fit the $\rho^0$ yield with a relativistic
Breit-Wigner function whose mean and width are fixed to the
PDG values~\cite{PDG2002} to estimate the branching fraction.

To study the color-suppressed decay mode $\bar{B^0} \to D^{0} \rho^0$,
we require $M^2_{D^0 \pi^+}>14.0$~GeV$^2/c^4$
to remove backgrounds from $D^{*+} \pi^-, D_2^{*+} \pi^-$ decays and other $D$ 
resonances. After this requirement, we clearly
see an excess at the $\rho^0$ mass in the $\pi^+ \pi^-$ invariant mass 
distribution (Fig.~4(a)).
The excess around 1.45~GeV/$c^2$ can be modeled by either a $\rho(1450)$ 
or an $f_0(1370)$ resonance; we cannot discriminate between 
these states, or alternative models of the
excess, based on the fit.
Events near 0.5~GeV/$c^2$ may come from the
$\sigma$~\cite{sigma} resonance. We extract the $\rho^0$ yield using 
a one dimensional likelihood fit. We use a model which 
includes one low mass and one high mass wide resonance. 
The masses and widths are fixed, and the amplitudes and
phases are free parameters in the fit.
The error from the fit therefore incorporates the error from
the relative phases of the interfering terms: this tends to
increase the error on the yield.
The background under the signal events is described 
reasonably well by data from the $\Delta E$ sideband 
(0.06 GeV $<\Delta E <0.20$ GeV) shown
as the hatched histogram in Fig.~4(a). We model this shape with 
a combination of phase space, a polynomial and a 
Breit-Wigner function, where the third term takes into account the
possible contribution of true $\rho^0$ in the background.
From the fit, we obtain 86 $\pm$ 30 signal events corresponding to
a branching fraction of ${\mathcal B}(\bar{B^0} \to D^0 \rho^0)$ 
= (2.9 $\pm$ 1.0 $\pm$ 0.4)$\times 10^{-4}$. 
The statistical significance of the signal, calculated as 
$\sqrt{-2\ln(L_{0}/L_{max})}$ where $L_{max}$ is the likelihood with 
the nominal yield and $L_0$ is the likelihood with the signal 
constrained to be zero, is 6.1$\sigma$. 
We find a strong correlation
between the amplitude of the $\rho^0$ component 
and its relative phase with respect to the higher mass resonance;
if the amplitudes and phases of the high and low
mass resonances are fixed at their obtained values, and the fit is
repeated, a $\rho^0$ yield of 86$\pm$24 events is obtained.
We have repeated the fit with a number 
of different models including vector and scalar resonances
at different masses and with different widths; the variation in the
central value of the $\rho^0$ yield  
is negligible compared to the error from our default fit. 
As a further cross-check, we examine the
the helicity angle ($\Theta_{\rho}$), defined as the 
angle in the $\rho^0$ rest frame between the direction of the $\pi^+$
and the $\rho^0$ direction in the $B$ rest frame, and
find it to be consistent with the expected shape \cite{helicity}.

To extract the branching fraction of  $\bar{B^0} \to D^{*0} \rho^0$
we require $M^2_{D^{*0} \pi^+} > $ 6.3~GeV$^2/c^4$, which
removes backgrounds coming from 
$\bar{B^0} \to D_2^{*}(2460)^+ \pi^-$.   
We fit the $\pi^+ \pi^-$ mass distribution using a relativistic Breit-Wigner function 
after fixing the background shape from the $\Delta E$ sideband (Fig.~4(b)). 
The $\rho^0$ event yield is 29 $\pm$ 8, 
however since the limited statistics prevent us from studying
possible interferences with other resonances, we cannot 
interpret this as evidence for $D^{*0} \rho^0$ and we set an 
upper limit of the branching fraction.
Assuming Gaussian statistics, we find
$\mathcal{B}(\bar{B}^0 \to D^{\ast 0} \rho^0) < 5.1 \times 10^{-4}$ at the 90\%
confidence level. We summarize our results in Table~\ref{BR}.

\begin{figure}[h]
\begin{center}
\hbox{\epsfig{file=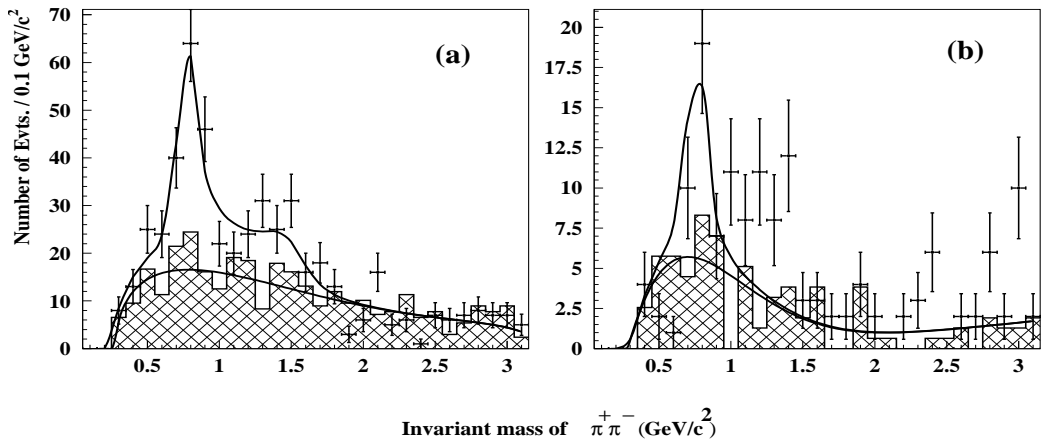,height=2.6in, width=5.5in}
}
\label{mpipimass_fig3}
\caption{
$M_{\pi^+ \pi^-} $ distribution from 
(a) $\bar{B^0} \to D^{0} \pi^+ \pi^- $ and (b) $\bar{B^0} \to D^{*0} \pi^+ \pi^- $ 
final states. The histogram represents
the data from the $\Delta E$ sideband and the fit to the histogram
is parameterized as described in the text.}
\end{center}
\end{figure}

\begin{table}[htb]
\caption{Summary of branching fraction results 
for $\bar{B^0} \to D^{(*)0} \pi^+ \pi^-$ and
$\bar{B^0} \to D^{(*)0} \rho^0$~\cite{footnote}.}
\label{BR}
\begin{tabular}{l c c c} 
\hline \hline
Mode & Efficiency (\%) & Branching Fraction ($\times 10^{-4}$) & Significance ($\sigma$)  \\
\hline
$\bar{B^0} \to D^0 \pi^+ \pi^-$ & 1.86 &  8.0 $\pm$ 0.6 $\pm$ 1.5  & 18.3 \\
$\bar{B^0} \to D^{*0} \pi^+ \pi^-$ &0.32 & 6.2 $\pm$ 1.2 $\pm$ 1.8 & 6.5\\
$\bar{B^0} \to D^{0} \rho^0 $ & 0.94 & 2.9 $\pm$ 1.0 $\pm$ 0.4 & 6.1 \\
$\bar{B^0} \to D^{*0} \rho^0 $ & 0.24 & $<$ 5.1 & - \\ \hline \hline
\end{tabular}
\end{table}
The following sources of systematic error are considered in our measurements:
(1) tracking efficiency (2.0\% per track - measured by
comparing the yield of the decay modes (i) $\eta \to \pi^0 \pi^+ \pi^-$ and
(ii) $\eta \to \gamma \gamma$); 
(2) particle identification efficiency for $\pi$ (4.3\%);
(3) $D^0$ reconstruction efficiency and decay branching fractions 
(12.7\% - measured by comparing the observed yield of
$B^- \to D^0 \pi^-$ events with the expected yield using known branching
fractions \cite{PDG2002});
(4) slow $\pi^0$ efficiency (10.7\% - measured from the ratio of
branching fractions of $B^- \to D^0 \pi^-$ and $B^- \to D^{*0} \pi^-$);
(5) $\Delta E$ signal parameterization (typically 8\%); 
(6) number of $B\bar{B}$ events (1.0\%) and 
(7) MC statistics (3--5\%).
As described previously, an additional systematic error due to
model dependence of the efficiency calculation
is added for $\bar{B^0} \to D^{(*)0} \pi^+ \pi^-$.
The total systematic error is obtained by
combining the different contributions in quadrature.

\section{Summary}

In summary, we report the first observation of the color-suppressed 
$\bar{B^0} \to D^0 \rho^0$ decays and measure the 
branching fraction for $\bar{B^0} \to D^{(*)0} \pi^+ \pi^-$.
Our measurement of ${\mathcal B}(\bar{B^{0}} \to D^0 \rho^{0})$ 
is higher than the factorization prediction of $0.7 \times 10^{-4}$
\cite{Neubert1}, thus continuing the 
trend mentioned in the introduction.
When we compare the branching fraction of $\bar{B}^0 \to D^0 \rho^0$ to our
previous measurement of the branching fraction of
$\bar{B}^0 \to D^0 \omega$~\cite{col_supp}, we obtain the
ratio ${\mathcal B}(\bar{B^0} \to D^0 \rho^0)$/${\mathcal B}(\bar{B^0} \to D^0 \omega)$ 
= 1.6 $\pm$ 0.8. The error includes both
statistical and systematic errors where the correlation of the
systematic errors  has been taken into account.
Future measurements with more statistics will allow precise
tests of the mechanisms involved in color-suppressed $B$ decays.

\section*{Acknowledgments}

We wish to thank the KEKB accelerator group for the excellent
operation of the KEKB accelerator.
We acknowledge support from the Ministry of Education,
Culture, Sports, Science, and Technology of Japan
and the Japan Society for the Promotion of Science;
the Australian Research Council
and the Australian Department of Industry, Science and Resources;
the National Science Foundation of China under contract No.~10175071;
the Department of Science and Technology of India;
the BK21 program of the Ministry of Education of Korea
and the CHEP SRC program of the Korea Science and Engineering Foundation;
the Polish State Committee for Scientific Research
under contract No.~2P03B 17017;
the Ministry of Science and Technology of the Russian Federation;
the Ministry of Education, Science and Sport of the Republic of Slovenia;
the National Science Council and the Ministry of Education of Taiwan;
and the U.S.\ Department of Energy.

\vfil


\begin{thebibliography}{99}
\bibitem{helicity}
Since $B^0$ and $D^0$ mesons are pseudoscalars, the $\rho^0$ mesons from
$\bar{B^0} \to D^0 \rho^0$ decays will be longitudinally polarized giving
a $\cos^{2}(\Theta_{\rho})$ distribution.
\bibitem{footnote}
The significance for $\bar{B^0} \to D^0 \pi^+ \pi^-$
is estimated by adding in quadrature the significances measured in each 
of the six Dalitz regions.
\end{thebibliography}
\end{document}